\documentclass[conference]{IEEEtran}
\IEEEoverridecommandlockouts
\usepackage{cite}
\usepackage{amsmath,amssymb,amsfonts}
\usepackage{amssymb}
\usepackage{algorithmic}
\usepackage{graphicx}
\usepackage{textcomp}
\usepackage{xcolor}
\usepackage{cite}
\usepackage{amsmath,amssymb,amsfonts}
\usepackage{algorithmic}
\usepackage{graphicx}
\usepackage{textcomp}
\usepackage{times}
\usepackage{xcolor}
\usepackage{float}
\usepackage{algorithm}
\usepackage[absolute]{textpos}
\usepackage{multirow}
\usepackage{mathtools}
\usepackage{mathtools, nccmath}
\usepackage{xcolor}
\newcommand{\blue}[1]{\textcolor{blue}{#1}}
\usepackage[font=small,skip=0pt]{caption}
\def\BibTeX{{\rm B\kern-.05em{\sc i\kern-.025em b}\kern-.08em
    T\kern-.1667em\lower.7ex\hbox{E}\kern-.125emX}}
\begin{document}

\begin{textblock}{18}(2,1)
\noindent\small \blue{This work has been submitted to the IEEE for possible publication. Copyright may be transferred without notice, after which this version may no longer be accessible.}
\end{textblock}

\title{Predictive Dynamic Scaling Multi-Slice-in-Slice-Connected Users for 5G System Resource Scheduling \\
}

\author{\IEEEauthorblockN{Sharvari Ravindran, Saptarshi Chaudhuri, Jyotsna Bapat, and Debabrata Das}
\IEEEauthorblockA{\textit{Networking and Communication Research Lab} \\
\textit{International Institute of Information Technology Bangalore (IIITB), India}\\
Email: Sharvari.R@iiitb.org, saptarshi.chauduri@iiitb.org, jbapat@iiitb.ac.in, ddas@iiitb.ac.in}
}

\maketitle

\begin{abstract}
Network slicing is an effective 5G concept for improved resource utilization and service scalability tailored to users (UEs) requirements. According to the standardization, 5G system should support UEs through specification of its heterogenous requirements such as, high data rates, traffic density, latency, reliability, UE density, system efficiency and service availability. These requirements are specified as Service Level Agreements (SLAs) between UEs and network operators resulting in increased interest to develop novel challenging mechanisms for improved interference-free resource efficient performances. An emerging concept that enables such demanding SLA assurances is Open-Radio Access Network (O-RAN). In this paper, we study a novel resource scheduling problem for UE services conditioned on other services in network slicing. To improve system performance, we propose that UEs are connected across network slices to serve several applications. UEs of similar service classes (defined by SLAs) are grouped to form optimized slice-in-slice category(s) within network slice(s). We propose novel Predictive \underline{D}ynamic Scaling \underline{M}ulti-\underline{U}E service specific \underline{S}ystem Resource \underline{O}ptimized Scheduling (DMUSO) algorithm(s). Multi-objective multi-constraint optimization problems (MOP) are formed to learn the dynamic system resource allocation and throughput for UE services conditioned on new services entering the network slice. An epsilon-constraint line search algorithm is presented to estimate UE service bandwidth. Using the theoretical models, DMUSO forms optimal slice-in-slice categories and estimates the maximum dynamic additional slice-in-slice categories, throughput served across network slices. Finally, compared to state-of-the-art literatures, DMUSO guarantees UEs SLAs with 4.4 and 7.5 times performance gains.
\end{abstract}


\section{Introduction}
In 5G, the concept of network slicing has been introduced for better resource utilization efficiency, flexibility and support of fast growing over the top (OTT) services. Network slice (or slice) is an independent dynamically created logical entity with a set of functions, services and system resources. Each network slice allows operators to provide services tailored to user equipments (UEs) requirements over a single Radio Access Network (RAN). According to the 3GPP standardization, 5G and Beyond should support UEs multiple needs such as, data rates, service availabilities, latency, reliability and traffic densities with resources availability. These UE specific capabilities form the basis for developing flexible and intelligent mechanisms to support heterogeneous services, multi-connectivity, on-demand service deployment and ensure guaranteed Service Level Agreement (SLAs). The advent of the Open-Radio Access Network (O-RAN) base station architecture with support of intelligent models achieves efficient system optimization. A UE entering the network gets associated with the slices in RAN and its services. To ensure guaranteed SLAs, the RAN slice scenario starts with UE's requirements [1]. The RAN is accordingly controlled by monitoring UEs long-term trends and patterns to achieve system resource allocation. However, owing to the surge in UE capacity, varying service data rates, fluctuating traffic and real-time channel variation,  the dynamic network slice management is critical and challenging. In view of this, new mechanisms need to be developed which requires allocating resources optimally to accommodate future time-varying service demands [3]. 
 \par
The relevant recent literatures of medium access control (MAC) scheduling and slicing in 5G have been reviewed. In [2], the authors have developed a resource scheduling mechanism to minimize power consumption in O-RAN slicing. The scheduling is achieved by mapping the slices to services and resources to slices, as a mixed integer optimization problem. The authors in [6] have
projected a RAN slicing method that flexibly allocates resources using deep reinforcement learning. The 
method ensures optimal allocation independent of the number
of slices, without overprovisioning resources. The work in [15]\textendash[17] have focused on priority-based \underline{r}esource \underline{a}llocation in \underline{n}etwork \underline{s}lices (RANS algorithm [15]) and \underline{r}esource \underline{o}ptimized \underline{s}cheduling \underline{s}trategy (ROSS algorithm [16]) for evaluating users priority depending on the application. The work in [4][13] have achieved resource allocation considering the SLA contracts, small-time scale network dynamics, network states and traffic data across slices. In [14], authors have formulated a fairness-based \underline{m}ulti-resource framework based on \underline{O}rdered \underline{W}eighted \underline{A}verage (MOWA algorithm) to satisfy users’ requirements, where the relation between resources is linear. However, the work does not consider SLA constraints to achieve resource scheduling. 
\par
Most of the prior work have relied on pre-negotiated SLAs, which monitor the responsibilities of UEs and define the fractional resource allocation. A reasonable approach to satisfy SLAs is triggering a reconfiguration process to allocate additional resources to UEs due to the uncertain demands. However, above existing methods of system resource scheduling will fail to handle,
\begin{enumerate}
\item Optimal resource allocation for UEs to prevent over-provisioning the system due to UEs fluctuating demands.
\item Network scalability support for UEs and services. The single hardware of a deployed O-RAN base station (gNodeB) of network slice must be scalable and intelligent to accommodate UEs services owing to the heterogeneous complexity of wireless systems.
\item Analysing the channel interference across UEs with respect to optimal UE-specific system resource allocation.
\item With the concept of dynamic infrastructure sharing [8], since the resources are pooled and co-existence of multiple UEs, an efficient and fair framework is needed. 
 \end{enumerate}
\begin{figure}
\begin{center}
\includegraphics[scale = 0.24]{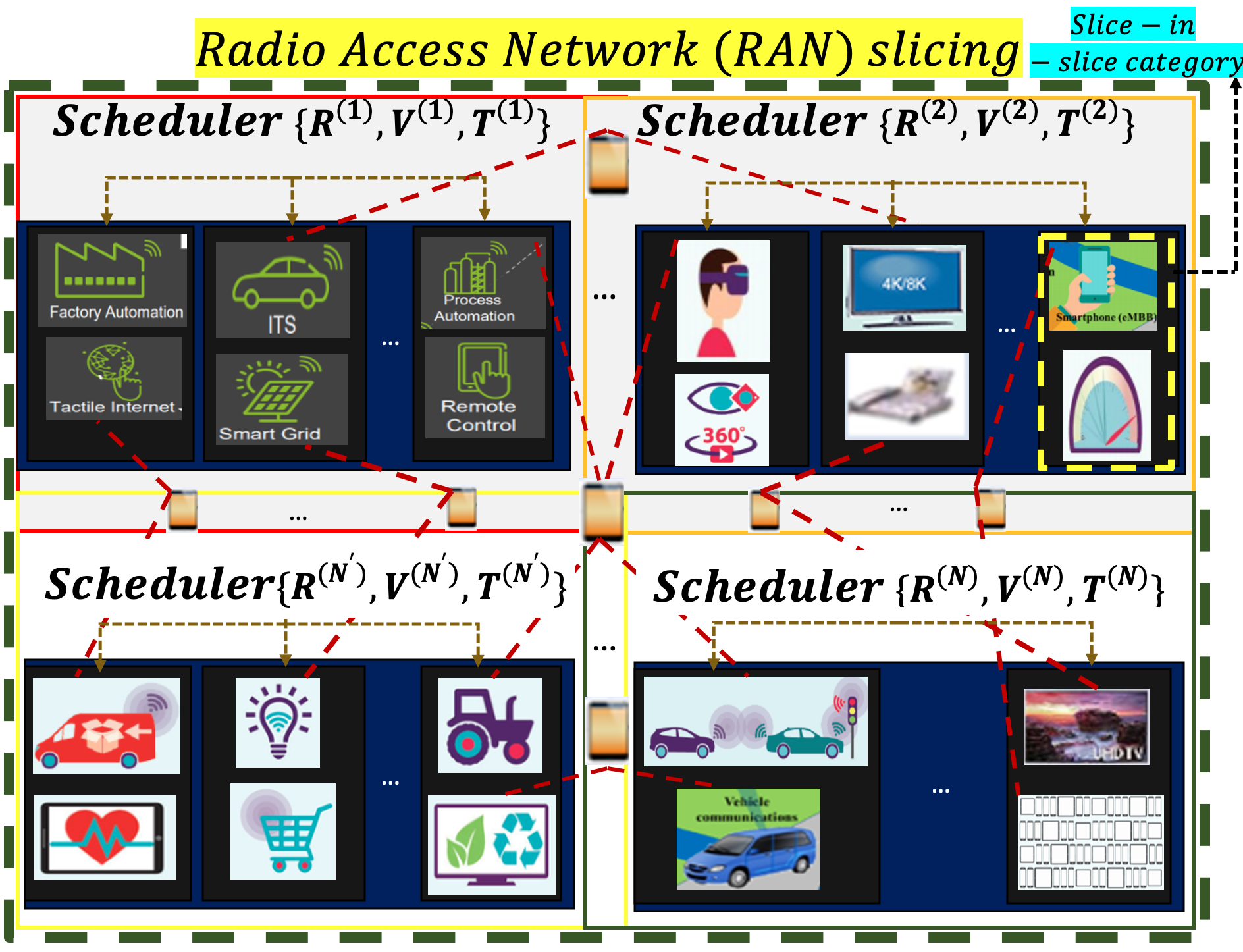}
\end{center}
\caption{Pictorial representation of multi-slice-in-slice-connected UEs}
\vspace{-5mm}
\end{figure}
\hspace{0.1cm}
Considering all the aforementioned challenges, ensuring quality by learning the service demands, data rates, resource, bandwidth, efficiency, transmission rates and channel related SLAs have to be yet addressed extensively for a UE-specific level in network slicing. In view of this, to improve system performance, we propose the concept of multi-slice-in-slice-connected UE(s) to serve applications. Fig. 1 depicts a novel representation of multi-slice-in-slice-connected UEs across network slices $1,…,N$. Single and Multiple UEs gets associated with network slices and serves several different applications. The UE(s) services having similar SLAs ranges (defined by service classes) are grouped to form multiple slice-in-slice categories within each network slice. These slice-in-slice categories operate within the resource envelope of its encompassing slice. We summarize our novel contributions,
\begin{enumerate}
\item A predictive dynamic scaling multi-slice-in-slice-connected UE(s) services for System resource Optimized Scheduling (DMUSO) algorithm across time intervals. 
\item An algorithm for formation of optimized slice-in- slice service categories across network slice(s).
\item An algorithm to dynamically learn interference-aware UE service system resources and throughput conditioned on new services. The predicted service throughput estimates the maximum additional slice-in-slice categories and throughput for network slice(s). 
\item An algorithm of learning the Pareto optimal bandwidth of UE(s) service(s) for resource scheduling.
\item Extensive system-level simulations using our novel analytical models to validate DMUSO performance. We show that the resources are allocated optimally leading to improved system performance gains compared to state-of-the-art methodologies.
\end{enumerate}
\hspace{0.1cm}
The paper is structured as follows: Section II highlights the multi-slice-in-slice system parameters. Section III describes our proposed DMUSO algorithm and extensive analytical models. Section IV provides the simulation parameters. Section V describes the results and discussion. Finally, Section VI highlights the conclusions and future scope.

\section{Multi-Slice-in-Slice System Parameters}
This section describes the novel concepts of multi-slice-in-slice categories and the key implementation phases to achieve optimal and efficient system resource scheduling. \\
\textbf{Definition 1}:  \textit{A} $n^{th}$ \textit{multi-slice-in-slice category is an amalgamation of UE(s) similar SLAs, i.e., service class defined by} $\{\bigcup_{x = -\delta_{T^{(n)}}}^{\delta_{T^{(n)}}} T^{(n)} + x, \bigcup_{x = -\delta_{E^{(n)}}}^{\delta_{E^{(n)}}} E^{(n)} + y\}$, \textit{where} $T^{(n)}, E^{(n)}$ \textit{are the throughput, spectral efficiency} (section III) \textit{and} $\delta_{T^{(n)}}$, $\delta_{E^{(n)}}$ \textit{are its respective variances}. In the O-RAN architecture [10], our work of scheduling focuses on interaction between MAC Open distributed unit (O-DU)-Low-PHY [7] Open Radio unit (O-RU) for UE specific channel interference aware resource allocation. The key implementation phases are,   \\
1) \textit{Model capability query} comprising of information for model training, such as system resources, network slicing infrastructure for implementing models and QoS-SLA constraint  (Section III). In the RAN, the system resources $\forall$ network slice(s) $N$ is defined as; radio ($R^{(N)}$), virtualized ($V^{(N)}$) and transport ($T^{(N)}$) (shown in Fig. 1), in vector form, 
\begin{equation}
\begin{split}
\{R^{(N)}, V^{(N)}, T^{(N)}\} & = \{|R_{1}^{(N)},..., R_{S}^{(N)}|, |V_{1}^{(N)},...,\\ & V_{S}^{(N)}|, |T_{1}^{(N)},..., T_{S}^{(N)}|\}
\end{split}
\end{equation}
where $R_{S}^{(N)}, V_{S}^{(N)}$, $T_{S}^{(N)}$ are the resources for $S^{th}$ slice-in-slice category, decomposed into $|R_{1,S}^{(N)}, ..., R_{U(S),S}^{(N)}|$, $|V_{1,S}^{(N)}, ..., V_{U(S),S}^{(N)}|$, $|T_{1,S}^{(N)}, ..., T_{U(S),S}^{(N)}|$ for UE(s) service(s) $1,...,U(S)$. In this work, we achieve \textit{Scheduling} with respect to \textit{Radio resources} (S-RS) and \textit{Transport resources} (S-TS).\\
2) \textit{Model selection and training} at every transmission time interval (TTI) of 1 msec. To achieve scheduling, we propose, \\
\textbf{Definition 2}: $\{R_{U(S),S}^{(N)}, constant \; V_{U(S),S}^{(N)}, T_{U(S),S}^{(N)}\}$ \textit{system resources} \textit{is learnt as a function of multiple network parameters} $\rightarrow$ \textit{bandwidth} ($B$), \textit{user throughput} ($T_U$), \textit{spectral efficiency} ($S_{e}$), \textit{interference} ($I_{a}$) and \textit{channel variation} ($CQ_v$). \\
3) \textit{Model deployment and inference}. 
Depending on the SLAs, scheduler maps a dynamic fraction ($\beta(.)$) of system resources for UE service(s). The resource allocation is formulated as a 2-tuple $(D_{m+k}^{(N)} =  \sum_{i=1}^{U(m+k)} \beta (R_{i,m+k}^{(N)}, T_{i,S}^{(N)}), \{R^{(N)}, T^{(N)}\}= \sum_{m+k=1}^{S} D_{m+k}^{(N)})$ where $D_{m+k}^{(N)}$ is the demand of an $(m+k)^{th}$ slice-in-slice category (\textbf{Defn. 3}). Inference counters are further used for tracking UE services ($U(m+k)$) at a slice-in-slice-level and maximum slice-in-slice categories ($S$) served. \\
4) \textit{Model performance monitoring and updating} comprising of aggregation of system metrics across UE services. Depending on environment feedback (channel conditions, new UEs services entering the slice), the performance evaluation module informs the system to update the current model. Subsequently, model is re-trained and deployed for inference.

\section{Proposed Prediction-based DMUSO Algorithm and Analytical Model}
This section is divided into three sub-sections. First sub- section focuses on the system model for UEs within network slice(s). Second sub-section focuses on dynamically learning the system resources and throughput of UE service(s). Third  sub-section focuses on estimating the Pareto optimal bandwidth of UE service(s) and complete DMUSO algorithm. 
   
\subsection{System model for multi-slice-in-slice-connected UE(s)}
   We consider the RAN slices $N_1,…,N_Q$ with $V_{1}^{(N_1)},...,V_{U}^{(N_1)}$ being the multi-slice-in-slice-connected UE services (or referred to as UE services) per $N_1$. If $U$ be the cardinality of set of UE services within $N_1$, $U = |V|$.  \\
\textbf{Definition 3}: Let $m$ and $k$ denote the existing and new slice-in-slice categories (formed) within a network slice. Let $i|_{i=1,2,...,U(m+k)} \in (m+k)$ (where $U(m+k)$: Total number of services) denote the service(s) $i$ in $(m+k)^{th}$ slice-in-slice category, whose capacity is $l_{m+k}(i)$. Then, capacity of network slice is ${\hat{l}} = \sum_{m+k=1}^{S} l_{m+k}(i)$.
Consider an orthogonal frequency domain multiplexing-based multiple input multiple output downlink transmission system (gNodeB to UE). Let $u$ denote the index of a UE. The received signal [12] for $u^{th}$ UE from gNodeB of a network slice $N$ is, 
\begin{equation}
y_{u}^{(N)} = \sqrt{t_{u}} h_{u}^{H}e_{u}x_{u} + \sum_{q (\neq i) \in N} \sqrt{t_{q}} h_{q}^{H} e_{q}x_{q} + n_{u}
\end{equation}
\hspace{0.1cm}
where $t_{u}$, $t_{q}$ are $u^{th}$, $q^{th}$ ($q \neq i$) UEs transmit powers allocated by O-RUs [2], $h_{u}$, $h_{q}$  are channel vectors between UE $u$, UEs $q$ and O-RUs, $e_{u}$, $e_{q}$ are unit norm beamforming [8] vectors from O-RUs to UEs, $x_{u}$, $x_{q}$ are information symbols for UE $u$, UEs $q$, $n_{u} \sim CN(0,N_{0}b_{i,m+k})$ is the receiver noise. Between UE $u$ and gNodeB, (2) $\Rightarrow$
\begin{equation}
\gamma_{i,m+k}^{(N)} = \frac{|h_{u}^{H} e_{u}|^2 t_{u}}{N_{0}b_{i,m+k} + \sum_{q (\neq i) \in N} |h_{q}^{H} e_{q}|^2 t_{q}}
\end{equation}
\hspace{0.1cm}
is the received signal to interference noise ratio for $u$, where $b_{i,m+k}$ is the estimated bandwidth (S-TS) of each $i \in u$ in $N$.

\subsection{Learning-based dynamic method of prediction of S-RSs and service throughput for UE(s) services}
While achieving system resource scheduling, it is necessary to know how slice-in-slice categories are formed. The assumption is we deliberately create room for formation of a new ($k$) slice-in-slice category within a network slice. Let $u_{i,m+k}(t)$ denote the UEs $i^{th}$ service throughput [3], 
\begin{equation}
u_{i,m+k}(t) = \frac{f_{d}\eta_{i,m+k}}{\Delta t} 
\end{equation}
\hspace{0.1cm}
where $f_{d} = 2^{\mu}*15kHz*12$, $\mu$ is numerology [5], $\eta_{i,m+k}$ is the spectral efficiency and $\Delta t$ is TTI (1 msec). In this work, we consider the channel of UE(s) to be Rayleigh. Hence, using the concept of exponential utility function [3], $\eta_{i,m+k}$ is modeled as $e^{\beta(r_{i,m+k}s_{i,m+k})}$. $\beta$ is a constant, $r_{i,m+k}$ is the S-RS allocation for UE service and $s_{i,m+k}$ is the average SNR. We define the objective ($\mathcal{O}$) as maximization of throughput across UE services, such that sum of $r_{i,m+k}$ across UE services should be less than the maximum S-RSs per network slice ($r_{max}$). 
\begin{equation}
\begin{cases}
\mathcal{O}: max \sum_{m+k=1}^{S} \sum_{i=1}^{U(m+k)} u_{i,m+k} (t) \\
Subject \; to,
\mathbf{C}: \sum_{m+k=1}^{S} \sum_{i=1}^{U(m+k)} r_{i,m+k} \leq r_{max}
\end{cases}
\end{equation}
\hspace{0.1cm}
Generally, as SNR increases, throughput increases. Due to the Rayleigh channel of UEs, $s_{i,m+k} \sim \frac{1}{\sigma_{s}^2} e^{-\frac{x}{\sigma_{s}^2}}$ of mean $\sigma_{s}^2$. \\
\begin{algorithm}
\caption{Learnable S-RS and UE service throughput}\label{euclid}
\begin{algorithmic}[1]
	\STATE $i_{m+k} = i_{current}|_{(m+k)_{current}} \leftarrow$ $i^{th}$ service $\in$ $(m+k)$ 
	\FOR {$k \neq 0$}
	\STATE $\{\lambda_{1}^{'}, ST^{'}\} \leftarrow \{\lambda_{1}, ST\}_{i=U(m+k),m+k=M}$
	\STATE $Go \; to \; Step \; 6|_{m = M, \; ST = ST^{'}, \lambda_1 = \lambda_{1}^{'} }$
	\ENDFOR
	\STATE $c_{M+\Delta}(t) = \sum_{m+k=1}^{S} \sum_{i=1}^{U(m+k)} b_{i,m+k} log(1 + \gamma_{i,m+k})$
	\STATE $i_{m+k}^{'} \leftarrow (1_{1},...i_{m+k})$ (\% Services)
	\FOR {$i_{m+k}^{'}$}
		\STATE $\{\lambda_1, \lambda_2\}_{i_{m+k}^{'}} \leftarrow Eqn. \; (6)_{i_{m+k}^{'}}$ 
		\STATE $\{FT, ST\}_{i_{m+k}^{'}} \leftarrow Eqn. \; (8)_{i_{m+k}^{'}}$
		\STATE $\{r_{i,m+k}\}_{i_{m+k}^{'}} \leftarrow Eqn. \; (7)_{i_{m+k}^{'}}|_{Step \; 9, 10}$
		\STATE $\{u_{i,m+k}\}_{i_{m+k}^{'}} \leftarrow \{\frac{f_{d} e^{\beta r_{i,m+k} s_{i,m+k}}}{\Delta t}\}_{\{r_{i,m+k}\}_{i_{m+k}^{'}}}$
	\ENDFOR
\end{algorithmic}
\end{algorithm}
\newline
\textbf{Lemma 1}: \textit{The PDF of SNR for existing $M$ slice-in-slice categories is} $\frac{x^{U(M)-1} e^{- \frac{x}{\sigma_{s}^{2}}}}{\gamma(U(M)) (\sigma_{s}^2)^{U(M)}}$, \textit{where} $U(M)$ \textit{are UEs across} $M$ [Appendix A [18]]. Suppose a new slice-in-slice category $k$ is formed. The network is benefited with a throughput increment, but with the cost of more $r_{i,m+k}$ and interference. \\
\textbf{Lemma 2}: \textit{The probability of formation of new slice-in-slice category conditioned on $M$ categories per network slice is $ \frac{e^{-\lambda_{2}} (\lambda_{2})^{m+k} m!}{e^{-\lambda_{1}} (\lambda_{1})^{m} (m+k)!}$}, \textit{where} $\{\lambda_{1}, \lambda_{2}\} \Rightarrow$ [See Appendix B [18]]
\begin{equation}
\begin{split}
\frac{f_{d}}{\Delta t}\{\sum_{m=1}^{M} m \frac{\sum_{i=1}^{U(m)} e^{\beta r_{i,m} s_{i,m}}}{c_{M}(t)} & , \sum_{m+k=M+1}^{M+\Delta} m+k \\ & \frac{\sum_{i=1}^{U(m+k)} e^{\beta r_{i,m+k} s_{i,m+k}}}{c_{M+\Delta} (t)}
\end{split}
\end{equation}
\hspace{0.1cm}1
where $c_{M}(t) = \sum_{m=1}^{M} \sum_{i=1}^{U(m)} b_{i,m+k} log(1 + \gamma_{i,m+k})$, $c_{M+\Delta}(t) = \sum_{m+k=M+1}^{M+\Delta} \sum_{i=1}^{U(m+k)} b_{i,m+k} log(1 + \gamma_{i,m+k})$ are cell throughputs (sum of UEs service throughput (based on [12])) of $m$, $m+k$ slice-in-slice categories, where $max(k) = \Delta$. Then, total PDF across all slice-in-slice categories is the sum of $m$ slice-in-slice category PDF (Lemma 1) and probability of formation of new slice-in-slice category (Lemma 2). To learn $r_{i,m+k}$, we differentiate with known $s_{i,m+k}$, $\frac{\partial (\frac{x^{U(M)-1} e^{- \frac{x}{\sigma_{s}^2}}}{\gamma(U(M)) (\sigma_{s}^2)^{U(M)}})}{\partial s_{i,m+k}} + \frac{\partial \frac{e^{-\lambda_{2}} (\lambda_{2})^{m+k} m!}{e^{-\lambda_{1}} (\lambda_{1})^{m} (m+k)!}}{\partial s_{i,m+k}} = 0$ $\Rightarrow$
\begin{equation}
\begin{split}
& (\frac{x^{U(M)-1}}{\gamma (U(M)){(\sigma_{s}^2)^{(U(M)-1)}}})_{k=0} = -FT_{k} \\ & (e^{-\lambda_{1}} ((\lambda_{1})^{m} (m+k)!)^{-1}) - ST_{k=0} (e^{-\lambda_{2}} (\lambda_{2})^{m+k} m!)
\end{split}
\end{equation}
\begin{algorithm}
\caption{Optimal S-TS}\label{euclid}
\begin{algorithmic}[1]
\STATE $X^{n} = b_{i,m+k} > 0, \epsilon, \epsilon^{'} > 0, c_{1} \in (0,1), c_{2} \in (c_{1},1)$
\FOR {$f_{2}(X^{n}) \leq  \epsilon$}
\STATE $d_{1}^{n} = -\triangledown f_{1}(X^{n}$): 
\STATE $Find \; \alpha_{1}^{n} \; along \; d_{1}^{n} \; s.t.$ 
\STATE {$f_{1} (X^{n} + \alpha_{1}^{n}d_{1}^{n}) \leq f_{1}(X^{n}) + c_{1}\alpha_{1}^{n}  \triangledown f_{1}^{k^{T}}d_{1}^{n}$} 
\STATE {$|\triangledown f_{1}(X^{n} + \alpha_{1}^{n}d_{1}^{n})d_{1}^{n}| \geq c_{2} \triangledown f_{1}^{k^{T}}d_{1}^{n}$}
\STATE $[n \leftarrow n + 1]$: $X^{n} = X^{n-1} + \alpha_{1}^{n}d_{1}^{n-1}$ 
\IF {$||f_{1}(X^{n}) - f_{1}(X^{n-1})|| < \epsilon^{'}$}
\STATE $b_{i,m+k} \leftarrow X^{n}|_{n_{optimum} = n_{current}}$
\STATE $\gamma_{i,m+k} \leftarrow Eqn. \; (3)_{b_{i,m+k} - Point \; 9}$
\ELSE
\STATE $Go \; to \; Step \; 2.$
\ENDIF
\ENDFOR
\end{algorithmic}
\end{algorithm}
$\{FT_{k}, ST_{k=0}\} = \frac{\partial (e^{-\lambda_{2}} (\lambda_{2})^{m+k} m!)}{\partial s_{i,m+k}} , \frac{\partial (e^{-\lambda_{1}} ((\lambda_{1})^{m} (m+k)!)^{-1})}{\partial s_{i,m+k}}$,
\begin{equation}
\begin{split}
& \Rightarrow \{\sum_{m+k=M+1}^{M+\Delta} \frac{m+k \sum_{i=1}^{U(m+k)} r_{i,m+k}s_{i,m+k}}{c_{M+\Delta}(t)} (e^{-\lambda_{2}} (m+k) \\ & \sum_{m+k=M+1}^{M+\Delta} m+k \frac{\sum_{i=1}^{U(m+k)} r_{i,m+k} s_{i,m+k}}{c_{M+\Delta}(t)})^{m+k-1} - \lambda_{2}^{m+k} \\ & e^{- \sum_{m+k=M+1}^{M+\Delta} m+k \frac{\sum_{i=1}^{U(m+k)} r_{i,m+k} s_{i,m+k}}{c_{M+\Delta}(t)}}), -(e^{-\lambda_{1}}(\lambda_{1}^{m}))^{-2} \sum_{m=1}^{M} \\ & m \frac{\sum_{i=1}^{U(m)} r_{i,m}}{c_{M}(t)} (m e^{-\lambda_{1}} (\sum_{m=1}^{M} m \frac{\sum_{i=1}^{U(m)} r_{i,m} s_{i,m}}{c_{M}(t)})^{m-1}) \\ &
- (\lambda_{1}) ^ {m} e^{- \sum_{m=1}^{M} m \frac{\sum_{i=1}^{U(m)} r_{i,m} s_{i,m}}{c_{M}(t)}}\}
\end{split}
\end{equation}
\hspace{0.1cm}
\textbf{Lemma 3}: \textit{The optimal weight} $\beta$ \textit{is learnt, such that} $\mathcal{O} (r_{i,m}^{*}) = 0$, where $r_{i,m}^{*} = r_{i,m}(\beta)$ [See Appendix C [18]].
To solve (8) for $r_{i,m+k}$, $u_{i,m+k}(t)$, we propose a low time complexity dynamic programming Algorithm 1. As new UE services enters the network slice, $r_{i,m+k}$, $u_{i,m+k}(t)$ for previous UE services is re-learnt by re-training the model(s), to accommodate more UE services. Using the models,  \\
$\mathbf{O1}$: \textit{Maximum additional slice-in-slice categories per network slice} ($\Delta$): Consider $\mathbf{C}$ in (5). Solving $\mathbf{C}$ for $U(m) = f_u$ (base number of UE services), yields $S$. Then, $\Delta = S - m$. \\
$\mathbf{O2}$: \textit{Maximum throughput of} $\Delta$ ($Thr(\Delta)$): Let $Thr(S)$ and $Thr(m)$ be the maximum cell throughput(s) for $S$ and $m$ respectively. Then, $Thr(\Delta) = Thr(S) - Thr(m) = \sum_{m+k=1}^{M+\Delta} \sum_{i=1}^{U(m+k)} u_{i,m+k} (t) - \sum_{m=1}^{M} \sum_{i=1}^{U(m)} u_{i,m} (t)$.

\subsection{Optimal S-TS for UE(s) services}
Multi-objective programming (MOP) provides several approaches to solve objectives leading to Pareto Optimality. For learning the S-TSs, i.e., $b_{i,m+k}$, we formulate the MOP
\begin{algorithm}
\caption{Predictive DMUSO}\label{euclid}
\begin{algorithmic}[1]
 \STATE $i \gets {1, 2,...,U}$ (UE services), $(N)$: Network slice \\ 
		\STATE $m \gets {1, 2,...,M}$ (existing slice-in-slice categories) \\ 
		\STATE $k \gets {0, 1, 2,...,\Delta}$ (additional slice-in-slice categories), 
		\STATE Initial: $m = 1, k = 0, Scheduling \; (S_{l}^{(N)}): \{...\}, l^{(N)}_{m+k} = \{...\}$ \\ 
		\STATE $t \gets {1, 2,...,T}$ (maximum simulation time) \\ 
		\FOR {$each \; TTI \; t \leq T$} 
\FOR {$(m + k)$ $\in$ $N$} 
\STATE $(m + k)^{th} \; category \leftarrow \{i = 1,2,...,U(m+k)\}$
	\FOR {$\forall \; i \in (m + k)$} 
	\STATE $S = (m + k)$
	\STATE $\{r^{(N)}_{i,m+k}(t), u^{(N)}_{i,m+k}(t)\} \; model \leftarrow Algorithm \; 1$ 
	\STATE $\{b^{(N)}_{i,m+k}, \gamma^{(N)}_{i,m+k}\}$ $\leftarrow$ $Algorithm \; 2$\\
	\STATE $\{r^{(N)}_{i,m+k}(t), u^{(N)}_{i,m+k}(t)\} \leftarrow Point \; 10|_{b^{(N)}_{i,m+k}}$ 
	\STATE $\eta_{i,m+k}^{(N)} \leftarrow \frac{u^{(N)}_{i,m+k}(t) (P_{c\{u\}} + t_{u})}{(P_{0\{u\}} + \phi_{u} t_{u})b^{(N)}_{i,m+k}}$ (\%Sec. III-C)
	\STATE $S^{(N)}_{l} \leftarrow \{(1,1)^{(N)}, r_{1,1}^{(N)}), ... \}$ (\%Updating) \\ 
	\ENDFOR
	\STATE $l_{m+k}^{(N)} \leftarrow U_{m+k}^{N} \leftarrow \sum_{i=1}^{U(m+k)} i$ (\%Updating) \\
	\FOR {$m^{(N)} \leq M^{(N)}$}
	\STATE $m^{(N)} \leftarrow (m+1)^{(N)}$. $Go \; to \; Step \; 8.$
	\ENDFOR
	\IF {$\sum_{m+k=M+1}^{M + \Delta} \sum_{i=1}^{U(m)} r^{(N)}_{i, m+k} \leq r^{(N)}_{max}$}
	\STATE $k \leftarrow k + 1$. $Go \; to \; Step \; 7.$
	\ELSE
	\STATE $[\Delta^{(N)} \leftarrow k_{max} \leftarrow k]: S^{(N)} = M^{(N)} + \Delta^{(N)}$
	\ENDIF
	\ENDFOR
	\STATE $\hat{l} \leftarrow U^{(N)} \leftarrow \sum_{m+k=1}^{M+\Delta} l_{m+k}^{(N)}$
	\STATE $Go \; to \; \textbf{Algorithm 4}.$
	\STATE $t \leftarrow t + 1. \; Go \; to \; Step \; 6.$
	\ENDFOR
\end{algorithmic}
\end{algorithm}
\begin{equation}
\{\mathbf{M1} : max_{b_{i,m+k}} \; \eta_{i,m+k}, \mathbf{M2} : max_{b_{i,m+k}} \; u_{i,m+k}(t)\}
\end{equation}
\textbf{(a) $\eta_{i,m+k}$}: From [11], $\eta_{i,m+k} = \frac{E_{i,m+k}}{(P_{c\{u\}} + t_{u})b_{i,m+k}}$, where $E_{i,m+k}$ and $P_{c\{u\}}$ are the energy efficiency and system circuit power. Inserting $E_{i,m+k} = \frac{u_{i,m+k}(t)}{(P_{0\{u\}} + \phi_{u} t_{u})}$ [9] yields 
$\eta_{i,m+k} = \frac{u_{i,m+k}(t)}{(P_{c\{u\}} + t_{u})(P_{0\{u\}} + \phi_{u} t_{u})b_{i,m+k}}$, where $P_{0\{u\}}, \phi_{u}$ are defined [9]. Substituting $u_{i,m+k}(t) = b_{i,m+k}log(1+\gamma_{i,m+k})$ [12] in $\eta_{i,m+k}$, (9): $\mathbf{M1}: max \; f_{1}$,
\begin{equation}
\mathbf{M1} : max \; \frac{log(1+\gamma_{i,m+k})}{(P_{0\{u\}} + \phi_{u} t_{u})(P_{c\{u\}} + t_{u})}
\end{equation}
\textbf{(b)} $u_{i,m+k}(t)$: From \textbf{(a)}, inserting $\eta_{i,m+k} = \frac{log(1+\gamma_{i,m+k})}{(P_{0\{u\}} + \phi_{u} t_{u})(P_{c\{u\}} + t_{u})}$ in (4), (9) yields: $\mathbf{M2}: max \; f_{2}$,
\begin{equation}
\mathbf{M2} : max \; \frac{f_{d} log(1+\gamma_{i,m+k})}{\Delta t (P_{0\{u\}} + \phi_{u} t_{u})(P_{c\{u\}} + t_{u})}
\end{equation}
\textbf{Lemma 4}: \textit{MOPs} $\mathbf{M1}$ and $\mathbf{M2}$ \textit{are convex functions of $b_{i,m+k}$} [See Appendix D [18]]. To estimate the optimal solution, we re-express $\mathbf{M1}$ and $\mathbf{M2}$ as, 
\begin{equation}
\begin{cases}
\mathbf{M1}: min \; {\frac{{(P_{0\{u\}} + \phi_{u} t_{u}) (P_{c\{u\}} + t_{u})}}{log(1+ \gamma_{i,m+k})}} \\
\mathbf{M2}: min \; \frac{\Delta t {(P_{0\{u\}} + \phi_{u} t_{u}) (P_{c\{u\}} + t_{u})}}{f_{d} log(1+\gamma_{i,m+k})}
\end{cases}
\end{equation}
\hspace{0.1cm}
To solve (12), we employ the $\epsilon-$constraint and line search optimization methods, explained in Algorithm 2. 
\begin{algorithm}
\caption{Slice-in-slice category optimal formations}\label{euclid}
\begin{algorithmic}[1]
\STATE $m+k = 1, i = 1, \delta_{T^{(n)}} = 0.05 Mbps, \delta_{E^{(n)}} = 0.05 bits/sec/Hz \; (user-defined \; \forall \; n), SS^{(N)} = \{...\}$
\FOR {$i^{(N)} \in l_{m+k}^{N} \leq \hat{l} \in S^{(N)}_{l}$}
\STATE $ Sn^{(N)} \leftarrow \{\bigcup_{x = -\delta_{T^{(n)}}}^{\delta_{T^{(n)}}} u^{(N)}_{i,m+k} + x, \bigcup_{y = -\delta_{E^{(n)}}}^{\delta_{E^{(n)}}} \eta^{(N)}_{i,m+k} + y\}$ (\%\textbf{Defn. 1}: Store in $SS^{(N)}$)
\STATE $Sn^{(N)} \leftarrow i^{(N)}$ $\leftarrow$ $S-RS \; allocation$
\STATE $i \leftarrow i + 1$
\IF {\{$u^{(N)}_{i,m+k}, \eta^{(N)}_{i,m+k}\} \in Sn^{(N)} \in SS^{(N)}$\}}
\STATE $Sn^{(N)} \leftarrow i^{(N)}$ $\leftarrow$ $S-RS \; allocation$
\ELSE
\STATE $n^{(N)} \leftarrow n^{(N)} + 1. \; Go \; to \; Step \; 3.$
\ENDIF
\ENDFOR
\STATE $(m+k)^{(N)} \leftarrow (m+k+1)^{(N)}. \; Go \; to \; Step \; 2.$
\end{algorithmic}
\end{algorithm}
\newline
\textbf{Theorem 1}: \textit{Let} $f_{1}$ \textit{be continuous and differentiable}, $X^{n} = b_{i,m}^{n}, d_{1}^{n} \in \mathbb{R}$, \textit{such that} $f_{1}(X^{n}  + \alpha_{1}^{n} d_{1}^{n})$ \textit{is bounded and} $\triangledown f_{1}^{n^{T}} d_{1}^{n} < 0$. \textit{Then, for} $c_{1} \in (0,1)$, $c_{2} \in (c_{1},1)$,
\begin{equation}
\begin{split}
\{\alpha_{1}^{n} & : f_{1} (X^{n} + \alpha_{1}^{n}d_{1}^{n}) \leq f_{1}(X^{n}) + c_{1}\alpha_{1}^{n}  \triangledown f_{1}^{k^{T}}d_{1}^{n} \\ &
\triangledown f_{1}(X^{n} + \alpha_{1}^{n}d_{1}^{n})d_{1}^{n} \geq c_{2} \triangledown f_{1}^{k^{T}}d_{1}^{n}\}
\end{split}
\end{equation}
$\alpha_{1}^{n}$ \textit{is an interior in a non-empty set} [See Appendix E [18]].  \\
\textbf{Theorem 2}: \textit{The solution} $\gamma_{i,m+k}$ \textit{as a function of} $b_{i,m+k}$ \textit{is weakly pareto optimal} (WPO) \textit{if} $f_{2} (\gamma_{i,m+k}) < \epsilon_{2}$ \textit{or} PO \textit{iff} $f_{2} (\gamma_{i,m+k}) = \epsilon_{2}$ [Appendix F [18]].  The complete DMUSO algorithm is shown in Algorithm 3 and 4.

\section{Simulation parameters}
This section describes the simulation parameters set for DMUSO. The simulation comprises of gNodeBs serving randomly distributed UE(s) services. For simulation we consider 4 network slices of bandwidth 40, 60, 70 and 80 MHz respectively. Each UE hosts 4 services across 4 network slices which are optimally grouped into slice-in-slice categories. The detailed system-level simulation parameters are in Table I.

 \begin{table}[htbp]
	\caption{Simulation parameters}
	\begin{center}
		\begin{tabular}{|p{3cm}|p{4.7cm}|}
			\hline
			\textbf{Parameter} & \textbf{Value} \\
			\hline
			Channel bandwidth & 100 MHz \\
			\hline
			Network slice(s) topology & Four network slices with $M = 6$, Each UE serves 4 applications across 4 network slices. Each slice-in-slice category hosts 5 UE(s) service(s) initially.  \\
			\hline
			Bandwidth part (BP: Maximum S-TSs) & 40 MHz ($N_{1}$), 60 MHz ($N_{2}$), 70 MHz ($N_{3}$), 80 MHz ($N_{4}$)\\
			\hline
			Maximum S-RSs  & 200 ($N_{1}$), 300 ($N_{2}$), 350 ($N_{3}$), 400 ($N_{4}$) \\
			\hline
			Total number of UEs associated with network slices & 205 ($N_{1}$), 335 ($N_{2}$), 385 ($N_{3}$), 455 ($N_{4}$) \\
			\hline
			Channel models & Extended pedestrian A model 5 Hz
			
			Extended vehicular A model 70 Hz
			
			Extended typical urban model 300 Hz
			
			Rayleigh fading model \\
			\hline
			UE mobility & 5 m/sec - 35 m/sec\\
			\hline
		        UE transmit power & 25 dBm \\
		         \hline
			Cell radius & 0.95 km \\
			\hline
			Modulation & QPSK for  - $\infty$ \textless SNR \textless 8 dB
			
			16-QAM  for  8 dB \textless SNR \textless 14 dB
			
			64-QAM  for 14 dB \textless SNR \textless $\infty$ \\
			\hline
			Simulation time & 100 seconds (10000 TTIs) \\
			\hline
			Simulators & Netsim and MATLAB\\
			\hline
		\end{tabular}
		\label{tab1}
	\end{center}
	\vspace{-8mm}
\end{table}

\section{Results and Discussions}
In this section, we analyse the optimum S-RS allocation performance across multi-slice-in-slice-connected UE services. For performance analysis, we compare DMUSO with relevant latest algorithms (Figs. 3 and 4), MOWA, ROSS and RANS referred in section I. To understand the S-RS performance, DMUSO estimates the optimal (or optimum) S-TS and interference effect across multi-slice-in-slice-connected UE(s). Fig. 2(a) shows the interference function ($log(1+SINR)$) vs optimal S-TSs across UE services. From Fig. 2 (a), $log(1+SINR)$ are weakly PO for range of S-TSs along X-axis. This is because along the Y-axis, the interference function increases as S-TSs increases, depicting strong PO. From the concept of strong PO [11], any strong PO is weakly PO $\Rightarrow$ $\nexists! \; b_{i,m+k}$ $\rightarrow$ $log(1+SINR)_{b_{i,m+k}} \leq log(1+SINR)_{b_{i,m+k}^{'}}$ $\rightarrow$ $b_{i,m+k}^{'}$ is PO. As the interference function decreases (Y-axis of Fig. 2(a)), at lower S-TSs, $log(1+SINR)_{b_{i,m+k}^{'}} \approx log(1+SINR)_{b_{i,m+k}}$ for $b_{i,m+k}^{'} \approx b_{i,m+k}$, depicting weakly PO.  DMUSO uses the S-TSs in learning the optimal S-RSs allocation, shown in Fig. 2(b). From Fig. 2(b), for $N_1,N_2,N_3,N_4$, as compared to the maximum S-RSs (Table I), DMUSO performs near (to maximum S-RSs) and optimized S-RS allocation across UE services. The optimal S-RSs estimates the UE service throughput. The sum of the throughputs across UE services over TTIs gives the average cell throughput. Fig. 3 shows the weighted mean average cell throughput (Mbps) across 10000 TTIs vs slice-in-slice categories served across $N_1,N_2,N_3,N_4$. From the X-Y plane of Fig. 3, the maximum slice-in-slice categories $S$ (including $M$ (Table I)) served by $N_1,N_2,N_3,N_4$ are \{41, 67, 77, 91\} owing to the maximum S-RSs resource constraint. As the number of slice-in-slice categories increases, average cell throughput increases across the network slices. A clearer view of the performance of initial few slice-in-slice categories can be seen in the zoom portion of Fig. 3. In addition, as compared to state-of-the-art algorithms, DMUSO ensures minimum and maximum gains of 4.5 and 7.5. The performance improvement is because DMUSO estimates the dynamic optimum S-RSs and throughput for each UE service, which is re-learnt to accommodate more UE services within network slices. We now analyze the average cell throughput as a function of SNR and S-RSs. 
\begin{figure}[h!]
\begin{center}
\includegraphics[scale = 0.085]{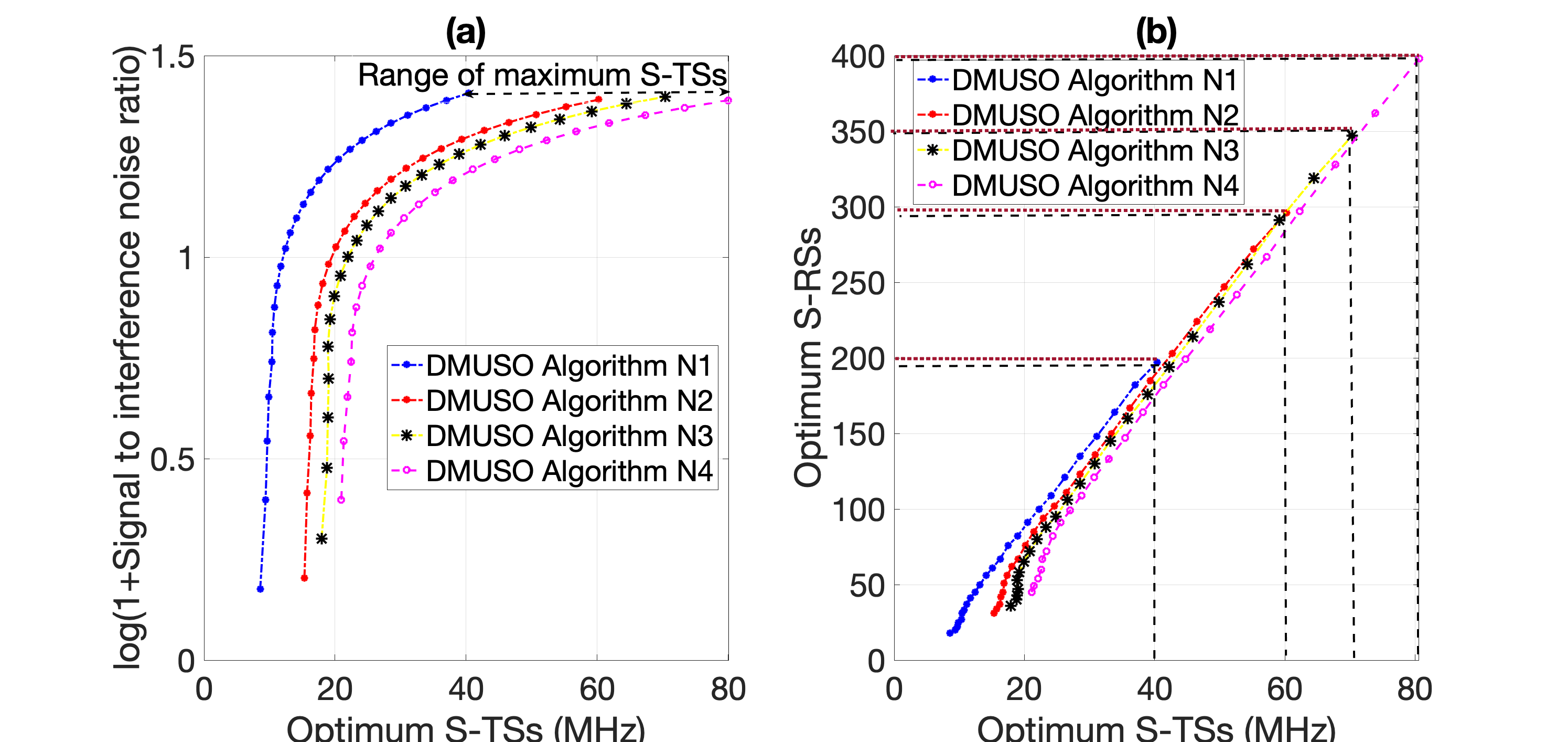}
\end{center}
\caption{(a) $log(1+SINR)$ vs Optimum S-TSs, (b) Optimum S-RSs vs S-TSs}
\vspace{-5mm}
\end{figure}
\begin{figure}[h!]
\begin{center}
\includegraphics[scale = 0.105]{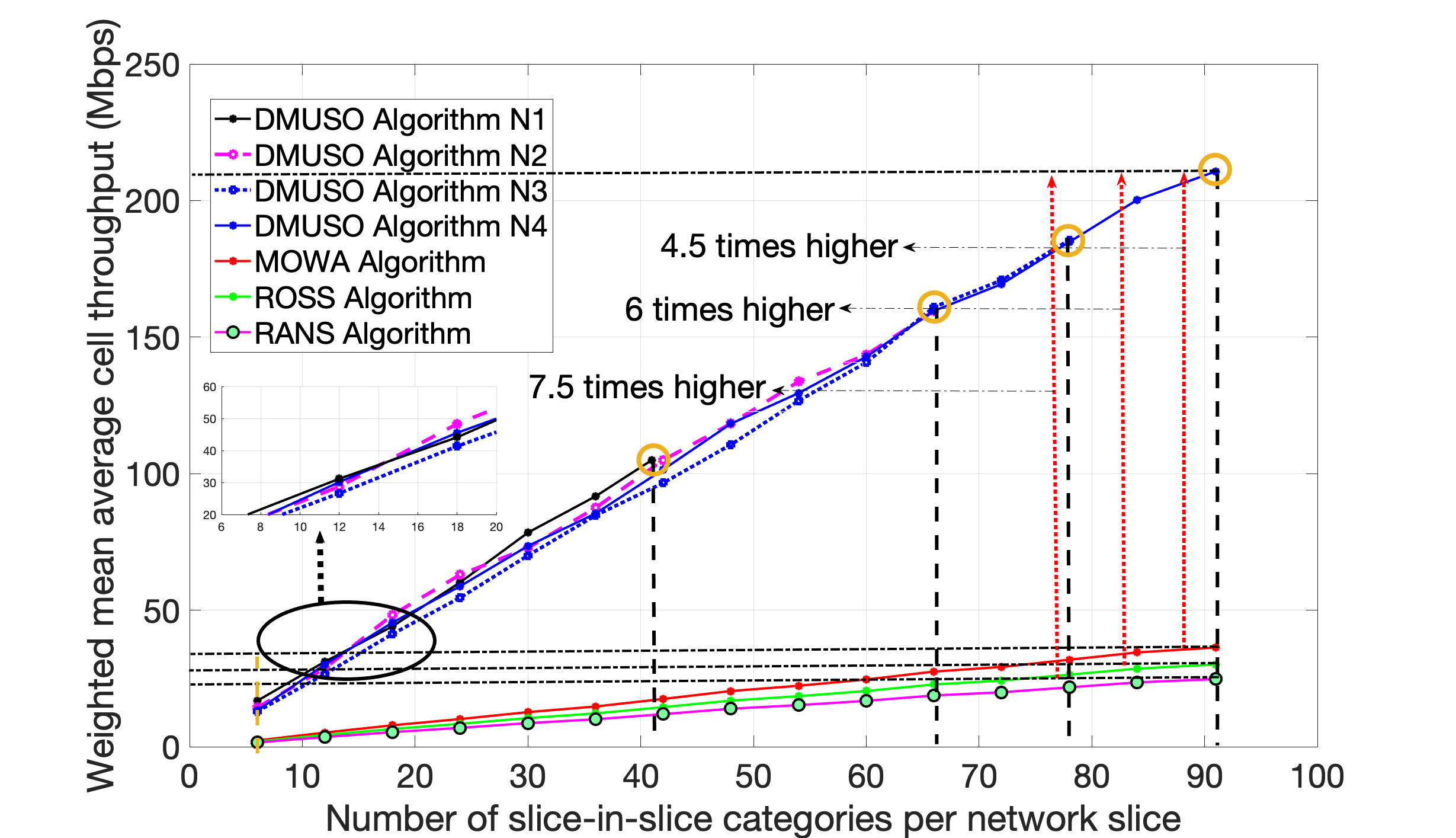}
\end{center}
\caption{Weighted mean average Cell throughput (Mbps) vs slice-in-slice categories across network slice(s)}
\vspace{-4mm}
\end{figure}
\begin{figure}[h!]
\begin{center}
\includegraphics[scale = 0.082]{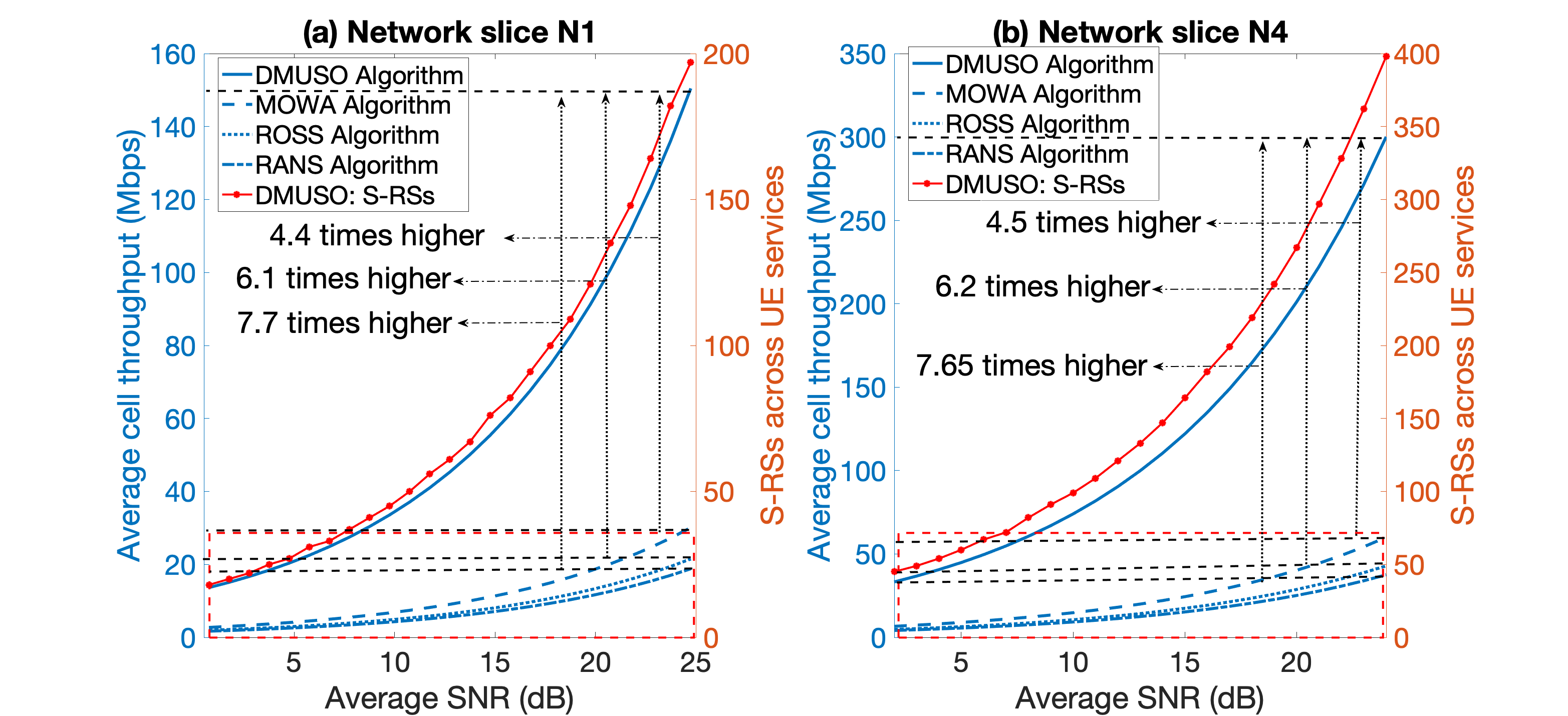}
\end{center}
\caption{Average Cell throughput (Mbps), S-RSs vs. Average SNR (dB)}
\vspace{-5mm}
\end{figure}
We plot the performances for the network slices having minimum and maximum BP, i.e., $N_{1}$ and $N_{4}$. Fig. 4 shows the average cell throughput (Mbps, in blue), average SNR (dB) and DMUSO S-RSs allocation (in red) for $N_{1}$ and $N_{4}$. From Fig. 4, the average cell throughput increases as average SNR increases. However, the average cell throughput for increasing SNR is better for DMUSO compared to state-of-the-art, owing to our learnable-assisted S-RS scheduling mechanism (right Y-axis of Figs. 4(a), 4(b)). The performance of state-of-the-art algorithms is comparitively lesser as the work (i) analyzes the relationship between resources, rather than satisfying UEs system demand (no SLA constraint(s)) [14], (ii) Prioritizing UEs and serving only certain UEs using specific criteria’s [15][16]. Hence, from Fig. 4, DMUSO achieves throughput gains of 4.4 and 7.65 times optimized and better compared to state-of-the-art. Thus, from the above extensive analysis and system-level simulations, we conclude that DMUSO guarantees UEs SLAs with improved performance gains across network slices. 

\section{Conclusions and Future scope}
To address the existing resource scheduling system challenges, we proposed the concepts of slice-in-slice categories and multi-slice-in-slice-connected UEs across network slices. We proposed novel predictive Dynamic Scaling Multi-slice-in-slice-connected UEs services for system resource optimized Scheduling (DMUSO) models and algorithms. Analysing the results, DMUSO achieves efficient and optimized system resource scheduling with significant performance gains of atleast 4.4 and 7.5 times compared to state-of-the-art algorithms. As part of future work, we intend to analyze the effect of varying UE mobility on resource scheduling across network slices.
   
\section*{Acknowledgment}

The authors would like to thank Advanced Communication System project sponsored by NM-ICPS scheme of Department of Science and Technology, Govt. of India.


\begin{thebibliography}{00}
\bibitem{b1} O-RAN WG 1 Slicing Architecture, v01.00, Technical Specfication. 
\bibitem{b1} M. Motalleb, V. Mansouri, S. Naghadeh, Joint Power Allocation and Network Slicing in an Open RAN System, in
arXiv:1911.01904, \textit{Computer Science, Networking and Internet Architecture}, Nov. 2019.
\bibitem{b1}  S. Chaudhuri, I. Baig, Debabrata Das, QoS aware Downlink Scheduler for a Carrier Aggregation LTE-Advance Network with Efficient Carrier Power Control, in  \textit{IEEE INFOCOM}, Bangalore, India, 2018.
\bibitem{b1}  L. Zhou, T. Zhang, J. Li, Y. Zhu, Radio Resource Allocation for RAN Slicing in Mobile Networks, Chongqing, China, \textit{IEEE/CIC International Conference on Communications in China}, Aug. 2020.
\bibitem{b1} 3GPP TS 38.300 v15.2.0 Release 15: 5G NR Overall description-2.
\bibitem{b1} Y. Saito, et. al., Flexible Resource Block Allocation to Multiple Slices for Radio Access Network Slicing Using Deep Reinforcement Learning, \textit{IEEE Access}, vol. 8, pp. 68183-68198, Apr. 2020.
\bibitem{b1} R. Stoica, G. Abreu, 6G: the Wireless Communications Network for Collaborative and AI Applications, in \textit{arXiv}, Apr. 2019.
\bibitem{b1}  O.U.Akgul, I. Malanchini, A. Capone, Dynamic Resource Trading in Sliced Mobile Networks, in \textit{IEEE Transactions on Network and Service
Management}, pp. 220-233, vol. 16, no. 1, Mar. 2019.
\bibitem{b1} S. Ravindran, S. Chaudhuri, J. Bapat, and Debabrata Das, EESO: Energy Efficient System-resource Optimization of Multi-Sub-Slice-Connected User in 5G RAN, in  \textit{IEEE International Conference on Electronics, Communication and Computing Technologies}, Bangalore, India, 2020.
\bibitem{b1} ORAN-WG2.AIML: AI/ML workflow description and requirements.
\bibitem{b1} L. Sboui et. al., A New Relation Between Energy Efficiency and Spectral Efficiency in Wireless
Communications Systems, in \textit{IEEE Wireless Communications}, vol. 26, no. 3, pp. 168-174, Jun. 2019.
\bibitem{b1} B. Matthiesen; O. Aydin; E. A. Jorswieck, Throughput and Energy-Efficient Network Slicing, in  \textit{22nd International ITG Workshop on Smart Antennas}, Bochum, Germany, 2018.
\bibitem{b1} M. Yan, G. Feng, J. Zhou, Y. Sun, Y. Liang, Intelligent Resource Scheduling for 5G Radio Access Network Slicing, in \textit{IEEE Transactions on Vehicular Technology}, vol. 68, no. 8, Aug. 2019, pp. 7691 - 7703.
\bibitem{b1}  F. Fossati, et. al., Multi-Resource Allocation for Network Slicing, in \textit{IEEE /ACM Transactions on Networking (Early Access)}, Mar. 2020.
\bibitem{b1} M. Dighriri et. al., Resource Allocation Scheme in 5G Network Slices, in \textit{32nd International Conference on Advanced Information Networking and Applications Workshops}, Krakow, Poland, May 2018.
\bibitem{b1} Y. Luo, X. Ye, X. Wang, H. Miao, T. Yang, An Optimal Scheduling Strategy for Wireless Resources based on Network Slicing, in \textit{IEEE 3rd Advanced Information Management, Communicates, Electronic and Automation Control Conference, Chongqing, China}, Oct. 2019.
\bibitem{b1} H. Ko, J. Lee, S. Pack, Priority-Based Dynamic Resource Allocation Scheme in Network Slicing, 2021 International Conference on Information Networking, Jeju Island, Korea (South), Jan. 2021. 
\bibitem{b1} Proofs of Lemmas and Theorems, Available [Online]: https://bit.ly/3dw9KPm
\end{thebibliography}
\end{document}